\documentclass[10pt,twocolumn]{paper}
\usepackage{epsfig, graphicx}
\usepackage[english]{babel}
\usepackage[margin=1.5cm]{geometry}

\title{\center \rm \bf Study of DMPS Monolayers on a Water Substrate with Laboratory X-ray Reflectometer}

\author{\small \rm Aleksey M. Tikhonov$^a$\/\thanks{tikhonov@kapitza.ras.ru}, Viktor E. Asadchikov$^b$, 
Yurii O. Volkov$^b$, Boris S. Roshchin$^b$, Yurii A. Ermakov$^c$\\
\small $^a$Kapitza Institute for Physical Problems, Russian Academy of Sciences, Moscow, Russia\\
\small $^b$Shubnikov Institute of Crystallography, Federal Research Center Crystallography and Photonics,\\
\small Russian Academy of Sciences, Moscow, Russia\\\small $^c$Frumkin Institute of Physical Chemistry and Electrochemistry, Russian Academy of Sciences,
Moscow, Russia}

\begin{document}
\maketitle
%\centerline{\today}

\abstract{ \it The molecular structure of dimyristoyl phosphatidylserine (DMPS) monolayers on a water substrate in different phase states has been investigated under normal conditions by X-ray reflectometry with a photon energy of $\sim 8$\,keV. According to the experimental data, the transition from a two-dimensional expanded liquid state to a solid gel state (liquid crystal) accompanied by the ordering of the hydrocarbon tails -C$_{14}$H$_{27}$ of the DMPS molecule occurs in the monolayer as the surface pressure rises. The monolayer thickness is $(20 \pm 3)$\,\AA{} and $(28 \pm 2)$\,\AA{} in the liquid and solid phases, respectively, with the deflection angle of the molecular tail axis from the normal to the surface in the gel phase being $26^\circ \pm 8^\circ$. At least a twofold decrease in the degree of hydration of the polar lipid groups also occurs under two-dimensional monolayer compression. The reflectometry data have been analyzed using two approaches: under the assumption about the presence of two layers with different electron densities in the monolayer and without any assumptions about the transverse surface structure. Both approaches demonstrate satisfactory agreement between themselves in describing the experimental results.}

\vspace{0.25in}
%\large

Phospholipids on a water surface form an insoluble monomolecular layer, a film that is a two-dimensional
thermodynamic system with parameters $(\Pi, T)$. Under certain conditions, the structure of the layer is
described by a symmetry axis perpendicular to the water–air interface [1, 2]. In particular, the Langmuir
monolayer formed by dimyristoyl phosphatidylserine (DMPS) molecules is such a system (Fig. 1).

Under normal conditions, the phase transition from a two-dimensional liquid to a gel structure (liquid
crystal) occurs in this film as the surface pressure $\Pi$ rises at constant temperature $T$ [3, 4]. However, simulations of such systems even using molecular dynamics methods leave the question about the
molecular nature of electrostatic effects in monolayers open. Considerable help in solving this question may
be expected from the use of direct methods of recording the structural changes in a monolayer. In particular, X-ray scattering was successfully used previously in [2, 5] to study the behavior of zwitterionic lipid
monolayers. In this paper we propose to use such a technique to study the monolayer structures of anionic
DMPS lipids with a pronounced phase transition. Indeed, under DMPS monolayer compression there
are two distinctly different regions of change in electric potential (potential drop in the lipid monolayer): a comparatively small and smooth change in potential in the liquid state of the monolayer gives way to its sharp increase ($\sim 200$\,mV) when the lipid passes into a solid gel phase. Various hypotheses [4], for example, a change in the hydration state of the polar phospholipid groups [5, 6], are proposed to explain this fact. We think that X-ray reflectometry data can be useful for testing these hypotheses. In this paper, based on our measurements of the X-ray reflectivity with a photon energy of $\sim 8$\,keV, we have reconstructed the electron density profile across the surface of a DMPS monolayer in its different phase states. Two approaches to analyzing the experimental data were used to extract the structural information: with a priori information ("model" approach) and without any assumptions about the transverse surface structure ("modelless" approach).

\begin{figure}
\hspace{0.1in}
\epsfig{file=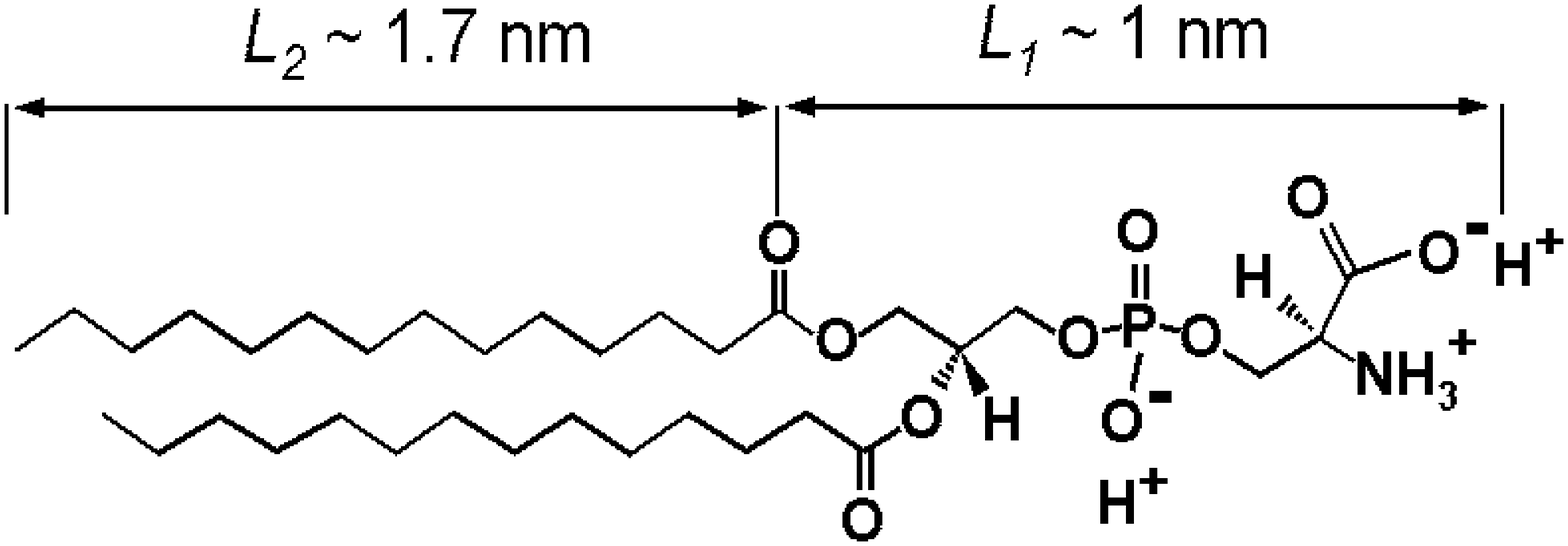, width=0.45\textwidth}

\small {\bf Figure 1.}  {\it Molecular structure of dimyristoyl phosphatidylserine (DMPS).}
\end{figure}

The samples of DMPS phospholipid monolayers were prepared and studied in an airtight cell with
X-ray-transparent windows in accordance with the technique described in [7, 8]. Some volume of a phospholipid solution with a concentration of 0.5\,mg/ml in a 5 : 1 chloroform–methanol mixture was spread with
a syringe over the surface of a liquid substrate (a KCl solution in deionized water with a concentration of
10\,mmol/L and pH=7) placed in a fluoroplastic dish with a diameter $D = 100$\,mm. The solution volume
required for our experiments was calculated for three values of the finite area $A$ per molecule chosen for different lipid phase states in the monolayer. The dependence of the surface pressure, $\Pi(A) = \gamma_0 - \gamma(A)$, measured on DMPS layers in [3, 4] is uniquely specified by $A$ (see Fig. 2). Here, the surface tension of pure water under normal conditions is $\gamma_0 = 72.5$\,mN/m, while $\gamma(A)$ is the surface tension in the presence of a lipid film. According to this dependence, at $A \approx 100$\,\AA$^2$ the monolayer is in an "expanded liquid" state I, while at $A \approx 50$\,\AA the monolayer is a two-dimensional inhomogeneous structure II and contains an equilibrium mixture of liquid and solid phase domains. Finally, at $A\approx 34$\,\AA$^2$ the DMPS monolayer is in a condensed state III that is usually defined as a liquid crystal [5, 9].

The transverse structure of the lipid monolayer was investigated by X-ray reflectometry on a versatile laboratory diffractometer with a movable emitter–detector system [10]. An X-ray tube with a copper anode is
used as the emitter. The K.1 line (photon energy $E =8048$\,eV, wavelength $\lambda = 1.5405 \pm 0.0001$\,\AA) is chosen from the tube emission spectrum using a single-crystal monochromator Si (111). The vertical and horizontal beam sizes are $\sim 0.1$ and $\sim 8$\,mm, respectively. The three-slit collimation system forms a probing X-ray beam with an angular width of $\sim 10^{–4}$\,rad in the plane of incidence. The angular resolution of the point detector $2\Delta\beta = 1.7 \times 10^{–3}$\,rad is determined by the entrance slit with a gap of 1\,mm at a distance of $\sim 570$\,mm from the sample center. Vacuum
paths with X-ray-transparent windows are used to reduce the absorption and scattering of emission in air.

Let ${\bf k}_{in}$ and ${\bf k}_{sc}$ be the wave vectors with an amplitude $k_0 = 2\pi/\lambda$ for the incident and scattered beams, respectively. It is convenient to introduce a coordinate system in which the origin $O$ lies at the center of the illumination region, the $xy$ plane coincides with the water boundary, the $x$ axis is perpendicular to the beam direction, and the $z$ axis is directed along the normal to the surface oppositely to the force of gravity (see the inset in Fig. 3). The scattering vector ${\bf q} = {\bf k}_{sc}-{\bf k}_{in}$ upon mirror reflection has only one nonzero component $q_z = 2 k_0 \sin\alpha$, where $\alpha$ is the grazing angle in a plane normal to the surface. The angle of total external reflection for a water surface $\alpha_c$ ($q_c = 2k_0\sin\alpha_c$) is fixed by the electron density in water $\rho_w=0.333$  {\it e-/}{\AA}$^3$, $\alpha_c\approx\lambda\sqrt{r_e\rho_w/\pi}$$\approx0.15^\circ$, where $r_e = 2.814 \times 10^{-5}$\,\AA is the classical electron radius.

\begin{figure}
\hspace{0.15in}
\epsfig{file=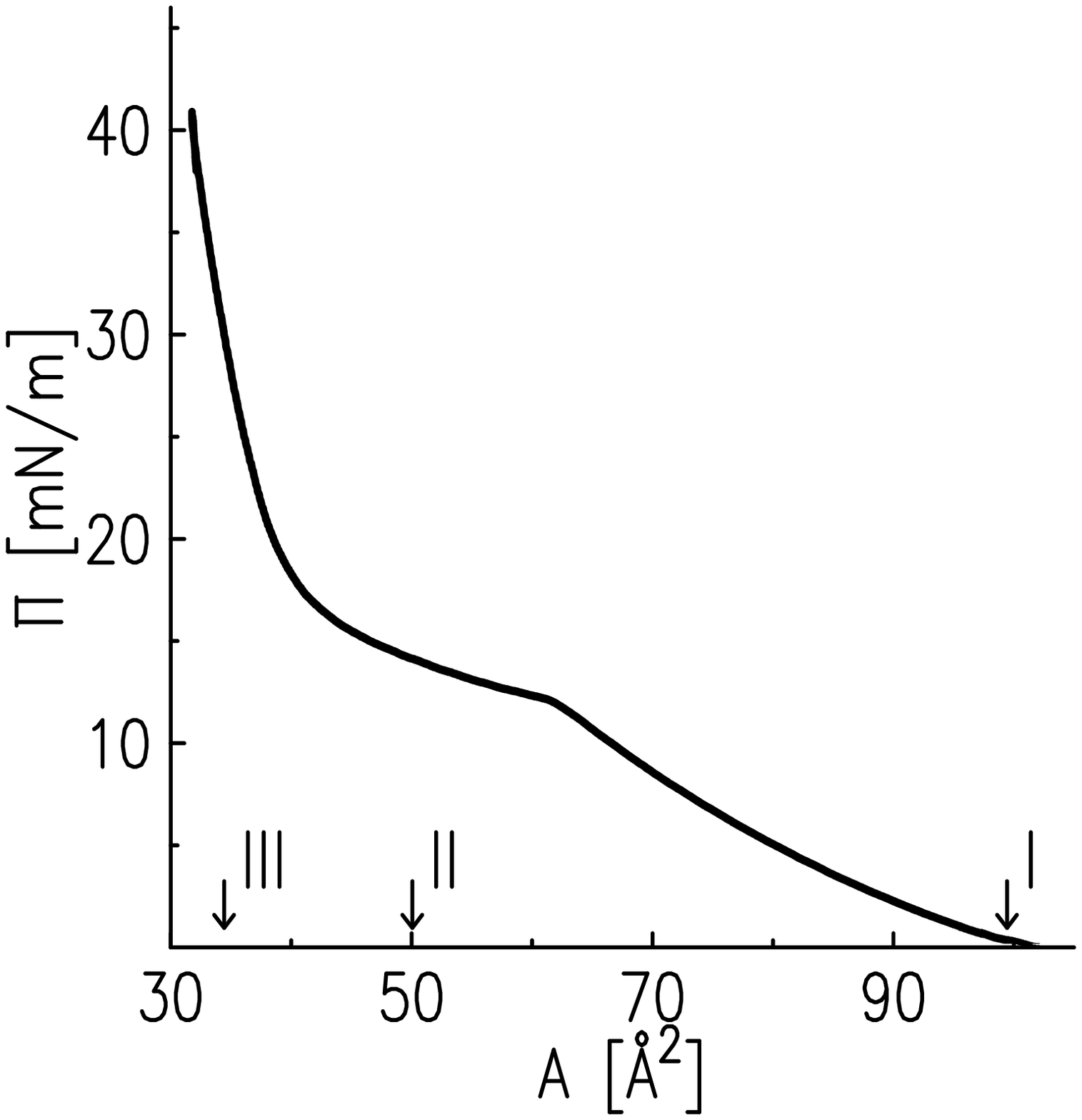, width=0.4\textwidth}

\small {\bf Figure 2.} {\it Measured dependence of the surface pressure on the area per molecule from [4] for a DMPS monolayer on the surface of a 10\,mM KCl and pH\,$\approx 7$ solution. The arrows indicate the areas chosen for X-ray reflectometry in the monolayer compression regions corresponding to an "expanded two-dimensional liquid" (I), an equilibrium mixture of liquid and solid phase domains (II), and a solid gel phase (III).}
\end{figure}

The diffractometer software allows a variable angular step, a detector slit width, and an exposure time to
be specified, which makes it possible to optimize the measurement of the reflectivity $R$ rapidly decreasing
with increasing $\alpha$. For a beam incident on the sample at an angle $\alpha$ the linear size of the illumination region along the beam is approximately $\propto 1/\sin\alpha$. As a consequence, at $\alpha \approx \alpha_c$ the beam section in the lateral $y$ direction (parallel to the sample surface) turns out to be appreciably larger than the sample diameter $D$, which leads to an incorrect determination of $R$. The correcting factor corresponding to the ratio of the total intensity of the direct beam to the intensity of its fraction falling within the sample surface is calculated before each measurement. The calculation of such a factor is similar to that in [11].

Figure 3 shows the dependence $R(q_z)$ measured in three independent experiments for a DMPS monolayer
on a water surface at various areas per molecule A near the phase transition: for $A \approx 100$ (state I), 50 (state II), and 34 (state III) \AA$^2$. At $q_z < q_c \approx 0.022$\,\AA$^{–1}$ the incident beam undergoes total external reflection, $R \approx 1$. Thus, the data for the reflectivity $R(q_z)$ collected on the diffractometer are comparable in spatial resolution $2\pi/q_z^{max} \approx 10$\,\AA{} ($q_z^{max} \approx 0.7$\,\AA$^{-1}$ is the maximum value of $q_z$ in our experiment) to the data obtained previously for various planar systems using synchrotron radiation [12-18].

\begin{figure}
\hspace{0.15in}
\epsfig{file=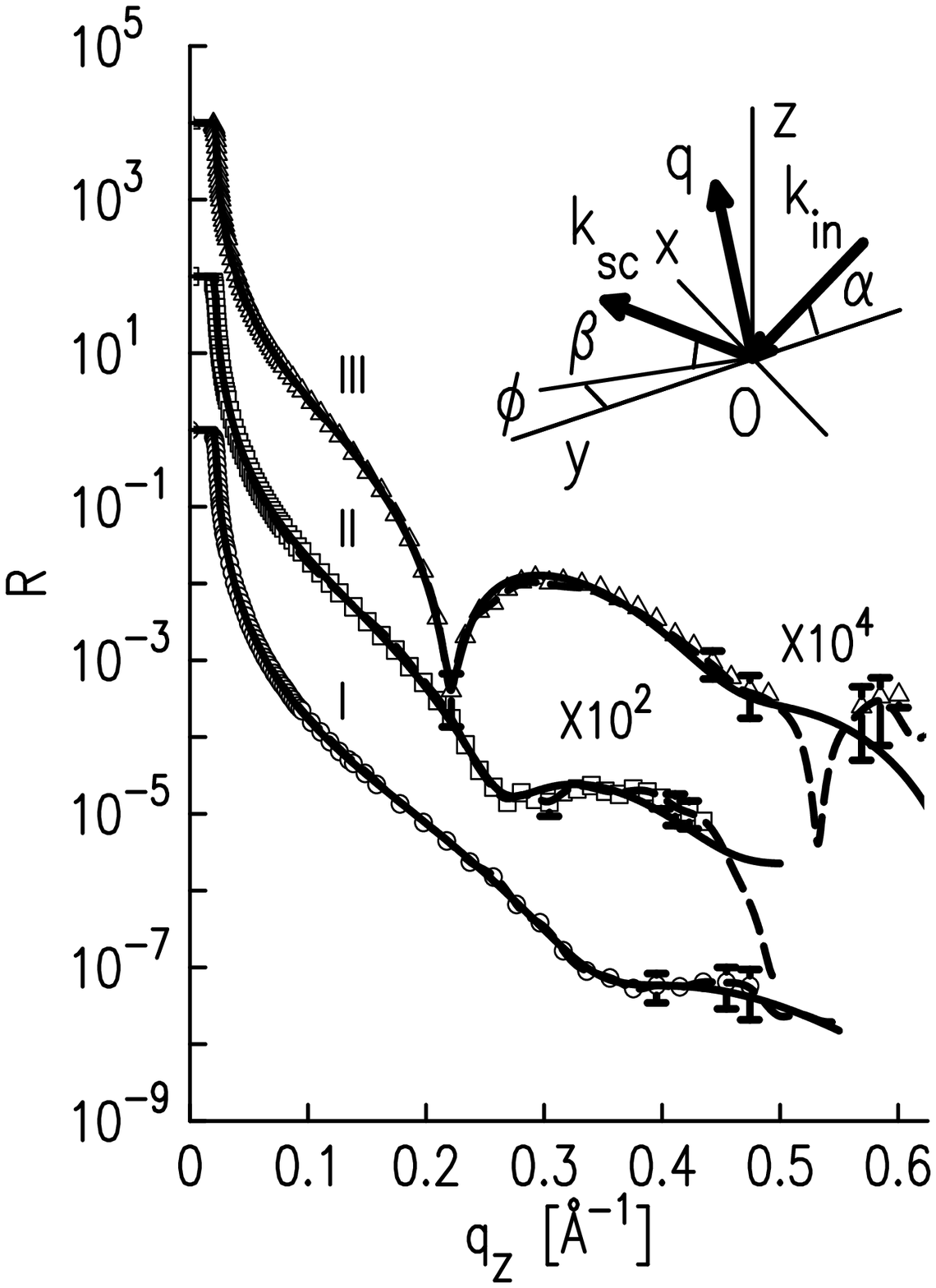, width=0.4\textwidth}

\small {\bf Figure 3.} {\it Dependence $R(q_z)$ for a DMPS monolayer on a water surface for various areas per molecule $A$: the circles, squares, and triangles are for $A = 100$ (state I), 50 (state II),
and 34 (state III) \AA$^2$, respectively. The solid lines indicate the two-layer monolayer model (model approach); the dashed lines indicate the result of the modelless approach when reconstructing the electron density profiles. Their difference is noticeable at large glancing angles. The inset: the scattering kinematics is described in a coordinate system in which the $xy$ plane coincides with the monolayer - water interface, the $Ox$ axis is perpendicular to the beam direction, and the $Oz$ axis is directed along the normal to the surface oppositely to the force of gravity.}
\end{figure}

The information about the surface structure in our experiment is averaged over a large illumination area
$A_0\approx 0.5$\,cm$^2$ and, therefore, the structure of the near-surface layer of our samples may be considered in the approximation of an ideal layer-inhomogeneous structure [19]. The electron density profile across the surface, $\rho(z)$, was reconstructed from the reflectometry data by invoking two different approaches.

The first approach is based on qualitative models with a minimum number of adjustable parameters
using a priori information about the molecular structure of a lipid film [17, 18]. For simplicity, below this approach is called the "model" one. In the distorted wave Born approximation (DWBA) the reflectivity for
a flat surface is [20]

\begin{equation}
R(q_z)=\left|\frac{\displaystyle q_z-q_z^t}{\displaystyle q_z+q_z^t}\right|^2\left|\Phi(\sqrt{q_zq_z^t})\right|^2,
\end{equation}
where $q_z^t=\sqrt{q_z^2 - q_c^2}$. Thus, interpreting the reflectometry data is reduced to finding some complex function of the structure factor $\Phi(q)$ that generally has the following form
\begin{equation}
\Phi(q)=\frac{1}{\rho_w}\int^{+\infty}_{-\infty}\left\langle\frac{d\rho(z)}{dz}\right\rangle e^{iqz} dz,
\end{equation}
where the electron density gradient is averaged over the area $A_0$.

In the case under consideration, for example, to achieve good agreement of the model curves with the
experimental data, it will suffice to divide the near-surface structure into two layers. In accordance with
the structure of the DMPS molecule, the first layer of thickness $L_1$ and electron density $\rho_1$ is formed by the polar phosphatidylserine groups, while the second layer of thickness $L_2$ and electron density $\rho_2$ is formed by the aliphatic tails -C$_{14}$H$_{27}$. Next, we construct the model profile $\rho(z)$ for the monolayer based on the error function ${\rm erf}(x)$ by assuming that all boundaries
between the layers and bulk phases have the same width $\sigma_0$ [21]:
\begin{equation}
\begin{array}{c}
\displaystyle
\rho=\frac{1}{2}\rho_{0}+\frac{1}{2}\sum_{j=0}^2(\rho_{j+1}-\rho_j)
{\rm erf}\left(\frac{l_j}{\sigma_0\sqrt{2}}\right),
\\ \\
\displaystyle
l_j=z+\sum_{n=0}^{j}L_n,
\\ \\
\displaystyle
{\rm erf}(x)=\frac{2}{\sqrt{\pi}}\int_0^x\exp(-y^2)dy,
\end{array}
\end{equation}
where $\rho_0 \equiv \rho_w$ is the electron density in water, $L_0 \equiv 0$ is the position of the water - polar group layer interface $(z=0)$, and $\rho_3\approx 0$ is the bulk electron density in air. Thus,
we have [22]

\begin{equation}
\begin{array}{c}
\displaystyle
R(q_z)=\frac{\exp(-\sigma^2_0q_zq_z^t)}{\rho^2_w}\left|\frac{\displaystyle q_z-q_z^t}{\displaystyle q_z+q_z^t}\right|^2
\\ \\
\displaystyle
\times
\left|\sum_{j=0}^{j=2}{(\rho_{j+1}-\rho_{j})\exp\left(-i\sqrt{q_zq_z^t}\sum_{n=0}^{j}L_n \right)}\right|^2.
\end{array}
\end{equation}

In our calculations the parameter $\sigma_0$ was fixed to be equal to the "capillary width"
$$
\sigma_{0}^2 = \frac{k_BT}{2\pi\gamma(A)}\ln\left(\frac{Q_{max}}{Q_{min}}\right)
$$
(where $k_B$ is the Boltzmann constant). The latter is specified by the short-wavelength limit in the spectrum of capillary waves $Q_{max} = 2\pi/a$ ($a\approx 10$\,\AA{} is the intermolecular distance in order of magnitude) and the angular resolution of the detector $Q_{min}=q_z^{max}\Delta\beta$. This method of allowance for the contribution of capillary waves to the observed structure on a liquid surface was proposed in [23-26] and turned out to be useful in interpreting numerous experiments [12, 14, 24-28]. Note that if $\sigma_0$ is used as an adjustable parameter, then its values for the systems being investigated lie in the range $2.8-3.1$\,\AA{} and coincide, within the error limits, with the calculated values of $\sigma_0 = 2.7-3.0$\,\AA{} from the dependence $\gamma(A)$, which also defines the compression diagram $\Pi(A)$.

The second approach is based on an extrapolation of the asymptotic behavior of the reflectivity curve
$R(q_z)$ to large $q_z$ without using any a priori assumptions about the transverse surface structure [29, 30]. This approach can be arbitrarily called the "modelless" one. In this approach the polarizability distribution in depth $\delta(z)$ is assumed to contain singular points $z_j$ at which the polarizability (or its $n$th derivative) changes abruptly:
$$
\Delta^{n}(z_j) \equiv  \frac{d^n\delta\left(z_j + 0\right)}{dz^n} - \frac{d^n\delta\left(z_j - 0\right)}{dz^n}.
$$

A combination of such singular points uniquely determines the asymptotic behavior of the amplitude
reflectivity $r(q_z)$ when $q_z \to \infty$ ($R(q_z) \equiv |r(q_z)|^2$). The arrangement of points $z_j$ can be determined from the experimental curve $R(q_z)$ measured in a limited range of values for $q_z$ using the procedure of a modified Fourier transform described in detail in [29]. Generally, there exist only two physically reasonable distributions $\delta(z)$ that simultaneously satisfy the experimental values of the reflectivity $R(q_z)$ and the specified combination of singular points $\Delta^n(z_j)$ in the polarizability profile and that differ only by the order of their arrangement relative to the substrate.

For each of the measured curves we found a pair of points with mutually opposite signs of the jumps in the
first derivative: the first corresponds to the air–sample interface, while the second corresponds presumably to the electron density maximum near the glycerin base of the polar group. The distance between them was
16.4, 23.5, and 25.4\,\AA{} for the films with $A$ equal to 100, 50, and 34\,\AA$^2$, respectively. The sought-for profile $\delta(z)$ was represented as a step function with fixed positions of the singular points $\Delta^1(z_j)$ and was divided into a large number $M$ $(M \approx 100)$ of thin layers: 
$\delta(z)= \sum_{m=1}^{M}{\Delta(z_m)H(z-z_m)}$,  where $H(z)$ is the Heaviside step function [31].
The reflectivity curve $R(q_z, \delta(z1), …,\delta(zM))$ for such a structure was calculated in accordance with Parratt's recurrence relations [32]. Thus, the polarizability profile was found by numerically
optimizing the residual between the experimental reflectivity curve and the calculated one regularized by
the smoothness condition for the sought-for profile and by the positions of the singular points using the
standard Levenberg–Marquardt algorithm [16, 33]. 

Finally, for weakly absorbing materials in the hard X-ray spectral range the electron density profile $\rho(z) \simeq \pi\delta(z)/(r_0\lambda^2)$ can be calculated from the reconstructed polarizability distribution in depth $\delta(z)$ [34].

Analysis of the data obtained confirms that the DMPS molecules are arranged on the water surface in
the form of a monolayer. In Fig. 4 the solid lines indicate the profiles $\rho(z)$ for the two-layer model (3), while the dashed lines indicate the profiles reconstructed within the modelless approach. The dependences $R(q_z)$ corresponding to these curves are represented in Fig. 3 by the solid and dashed lines. In Fig. 4 the difference between the two approaches becomes noticeable at large glancing angles, at which the experimental error increases significantly. To a first approximation, the measured and calculated curves presented in Figs. 3 and 4 show satisfactory agreement between the two approaches in describing the experimental results.

The model profile is characterized by four adjustable parameters (see the Table 1). As the surface pressure
rises, the electron density in the layer of polar phosphatidylserine heads directly in contact with the
water increases from $\rho_1\approx1.2\rho_w$ in state I to $\rho_1\approx1.4\rho_w$ in state III. At the same time, its thickness $L_1$ is virtually constant and lies in the range 10 - 12\,\AA{} for all states. In contrast, the thickness of the layer formed by the hydrocarbon chains increases noticeably from 
$L_2\approx 10$\,\AA{} (state I) to $L_2\approx 15$\,\AA{} (state III). Concurrently, the electron density also increases from $\rho_2\approx0.9\rho_w$ in the liquid phase to $\rho_2\approx0.95\rho_w$ in the solid phase. The total thickness $L$ of the monolayer is $L = (20 \pm 3)$\,\AA{} in state I and $L = (28 \pm 2)$\,\AA{} in state III. In intermediate state II the monolayer thickness is $L = (25 \pm 3)$\,\AA{}.

For the modelless description of the structures it will suffice to use only the first-order singular points,
because all experimental curves decrease approximately as $R(q_z)\sim q_z^{-6}$. The deviation from a strict
power law is apparently attributable to the scattering by sample surface roughnesses with an effective height $\sigma$. Its value can be estimated within the Nevot-Croce formalism from the requirements imposed on the
asymptotics
$$
R(q_z)q_z^6\exp\left(\sigma^2q_z\sqrt{q_z^2-4k_0^2\delta_+}\right)\to {\rm const}
$$
when $q_z \to \infty$, where $\delta_+ \approx 7.5\times 10^{–6}$ is the water polarizability for $\lambda \approx 1.54$\,\AA{} [19, 35]. Thus, we obtain $\sigma \approx 3.2$\,\AA, which agrees well with the calculated value of $\sigma_0 \approx 3$\,\AA{} given above. Note that the estimation of the integral roughness parameters from the reflectometry curves alone is highly ambiguous [36, 37]. For a more proper analysis of the statistical roughness properties of the sample being investigated, it is necessary to additionally invoke the angular distributions of diffuse scattering, for example, within the procedure described in [30].

\begin{figure}
\hspace{0.2in}
\epsfig{file=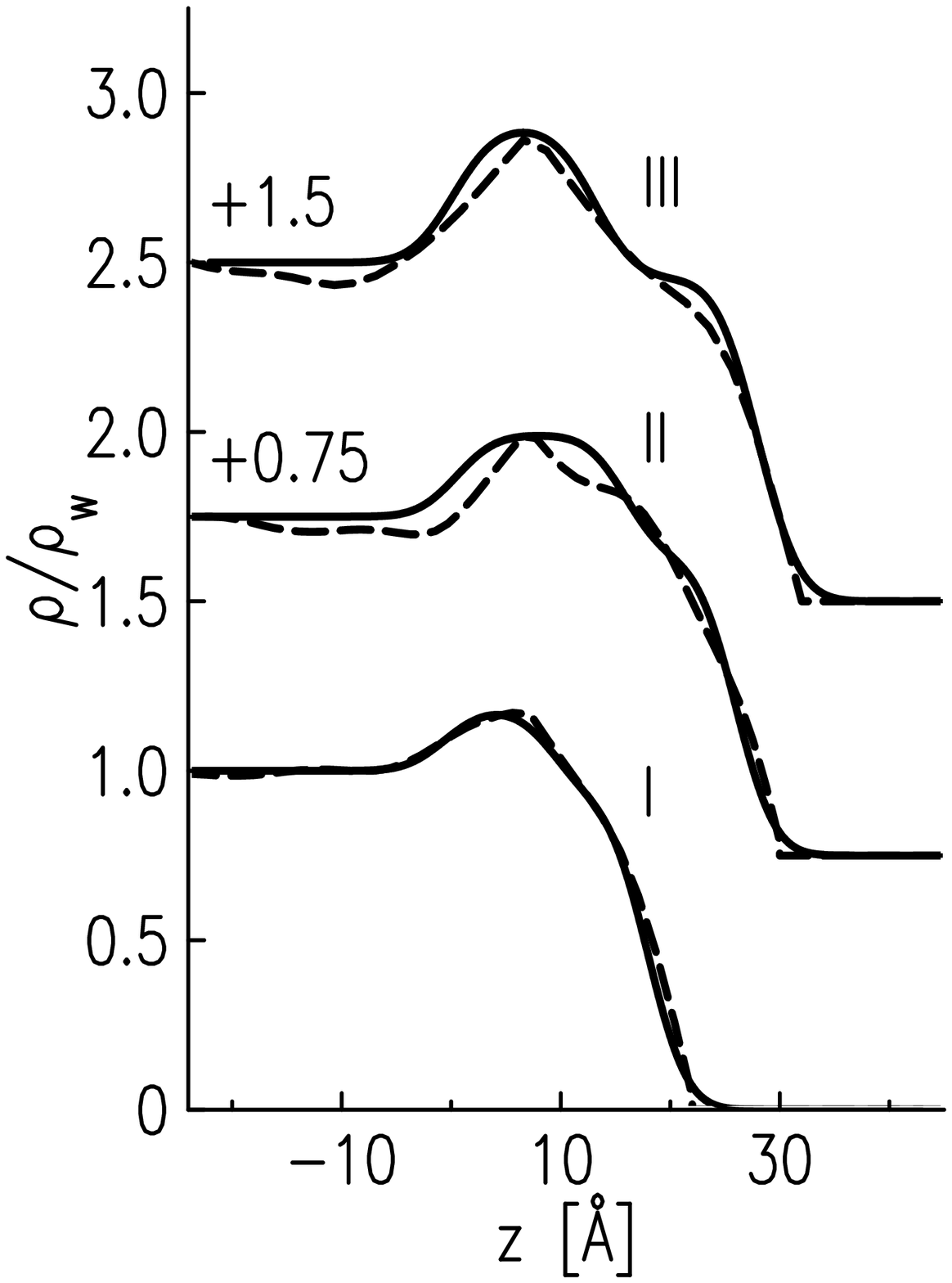, width=0.35\textwidth}

\small {\bf Figure 4.} {\it Electron density profiles normalized to the electron density
in water under normal conditions; the solid and dashed lines are for the model (see (3)) and modelless
approaches, respectively. The numbers near the curves specify their displacement along the vertical axis.}

\end{figure}

Next, given $\rho(z)$, the specific surface density $\Psi$ of structural units (ions, molecules, chemical groups) in a layer of thickness $d = z^{\prime\prime} - z^{\prime}$ can be estimated:

\begin{equation}
\Psi=\frac{1}{\Gamma}\int^{z^{\prime\prime}}_{z^{\prime}}\rho(z)dz,
\end{equation}
where $\Gamma$ is the number of electrons in one structural unit. For example, $\Gamma = 390$ for potassium salt C$_{34}$H$_{65}$NO$_{10}$4PK, $\Gamma_t/2 = 111$ for one -C$_{14}$H$_{27}$ chain, and $\Gamma_h = 168$ for the phosphatidylserine group.

For state I the thickness $L \approx 20$\,\AA{} and the distance between the singular points in the profile $\rho(z)$ of the modelless approach, approximately 16\,\AA, are appreciably smaller than the length of the lipid molecule, approximately 27\,\AA. This suggest that the hydrocarbon chains of molecules in the liquid phase of the lipid ($A\approx100$\,\AA$^2$) are disordered relative to the normal to the surface.

For state III ($A\approx34$\,\AA$^2$), the thickness of the second layer $L_2\approx15$\,\AA{} roughly corresponds to the calculated length of the hydrocarbon tails -C$_{14}$H$_{27}$, in the DMPS molecule, 
16.7\,{\AA} ($\approx  12\times 1.27$\,{\AA}(C-C) + 1.5\,{\AA}(-CH$_3$)). The density $\rho_2 \approx 0.95\rho_w$ and the area per hydrocarbon chain $A/2 \approx 17$\,\AA$^2$ correspond to one
of the crystalline phases of long-chain saturated hydrocarbons [38]. Thus, the deflection angle of the
molecular tail axis from the normal to the surface can be estimated:$\theta = \arccos(15/16.7)\approx 26^\circ$ ($26^\circ \pm 8^\circ$).

Note that the following imbalance in the number of electrons per structural unit is observed for all states.
For example, in state III the number of electrons accounted for by the polar part of the DMPS molecule
and the aliphatic tails is $\rho_1AL_1 \approx 206$\,{\it e$^-$} and $\rho_2AL_2 \approx 161$\,{\it e$^-$}, respectively. The excess electron density in the layer of heads is $A(\rho_1L_1 - \rho_2L_2\Gamma_h/\Gamma_t) \approx 84$\,{\it e$^-$} per DMPS molecule, which is equivalent to approximately eight H$_2$O molecules. Such a degree of hydration was reported previously for the gel phase of phospholipids in [5]. If the electron density is taken as a rough estimate of the degree of hydration, then it rises more than twofold as the area per molecule increases to $A\approx100$\,\AA$^2$ and is $\sim 20$ H$_2$O molecules per polar group.

According to the X-ray reflectometry data (Fig. 3), as the surface pressure rises, the phase transition from
an expanded two-dimensional liquid to a solid state becomes noticeable in the DMPS monolayer. The
main, and quite unexpected, result of our analysis of the experimental data is that the chosen two-layer
model of the structure (model approach) describes the electron density profile predicted within the modelless approach in a good approximation. This fact is illustrated by the parameters of the gel phase established
within these approaches and presented in Fig. 4 and the table. Both methods of analyzing the experimental curves give a pretty authentic idea of the behavior of the lipid monolayer as the lateral pressure changes. The model approach allows its important structural components to be identified, while the modelless approach allows one to independently confirm the correctness of the electron density distribution found and, thus, to reduce the ambiguity in interpreting the structural model.

Thus, we investigated the molecular structure of a dimyristoyl phosphatidylserine (DMPS) monolayer
on a water substrate in different phase states based on our X-ray reflectometry data. According to our analysis of the reflectivity curves, as the surface pressure rises, the transition from a two-dimensional expanded liquid state to a solid gel state accompanied by the ordering of the hydrocarbon tails -C$_{14}$H$_{27}$ occurs in the monolayer, while the thickness of the polar DMPS region remains virtually constant. The monolayer thickness is $(20 \pm 3)$ and $(28 \pm 2)$\,\AA{} in the liquid and solid phases, respectively. In the gel phase the deflection angle of the tail axis from the normal to the surface
is $26^\circ \pm 8^\circ$. At least a twofold decrease in the degree of hydration of the polar lipid groups occurs under two-dimensional monolayer compression. It is important to note that the decrease in the number of water molecules associated with the polar heads of lipids per se cannot lead to the positive change in electric potential observed in our experiments. Judging by the molecular dynamics data, the water and adsorbed cations are responsible for the positive changes in this potential [39]. Most likely, not the change in the number of water molecules and the degree of hydration but the orientation of their dipole moments and the adsorption of cations should be taken into account to explain the electrostatic effects in the monolayer. Detailed information about the molecules structures that are involved in such effects can be established by
molecular dynamics methods in combination with comprehensive experimental studies, including the
measurements of the reflectivity curves in a wide range of areas per molecule in the monolayer. We proved
that a quantitative analysis of the X-ray reflectometry data is fundamentally possible using several examples given above.

\small{
\vspace{5mm}
{\bf Table 1.} {\it Parameters of the electron density profiles $\rho(z)$ (see Fig. 4) for a liquid-crystal DMPS film at $A = 34$\,\AA$^2$. $L$ is the total thickness of the lipid layer, $L_1$ is the thickness of the layer of polar groups with an electron density $\rho_1$, $L_2$ is the thickness of the layer of hydrocarbon tails with a density $\rho_2$, and $\sigma_0$ is the width of the interlayer boundaries of the lipid monolayer. The electron density in water under normal conditions is $\rho_w=0.333$  {\it e$^-$/}{\AA}$^3$. The errors of the determination of the two-layer model parameters were obtained using the standard $\chi^2$ criterion at the confidence level of 0.9.}
\vspace{2mm}

\hspace{-8mm}
\begin{tabular}{|c|c|c|c|c|c|c|}
\hline
&&&&&\\
Approach &$L$({\AA})&$L_1$({\AA})&$L_2$({\AA})& $\rho_1 / \rho_w$ & $\rho_2 / \rho_w$ & $\sigma_0$({\AA})\\
&&&&&& \\
\hline
&&&&&& \\
Model & $28^{\pm1}$&$13^{\pm 1}$&$15^{\pm 1}$&$1.4^{\pm 0.1}$&$0.95^{\pm 0.02}$&$3.0$ \\
&&&&&& \\
Modelless &$\approx 30$&$\approx 13$&$\approx 17$& $\approx 1.3$&$\approx 0.95$&$\approx 3.2$\\
&&&&&& \\
\hline
\end{tabular}
}

\end{document}